\documentclass[12pt]{article} 
\usepackage{epsfig} 
\usepackage{latexsym} 
\usepackage{axodraw} 
\begin{document}

\begin{titlepage}

\begin{flushright}

\end{flushright}

\baselineskip 24pt

\begin{center}

{\Large {\bf Neutrinoless Double Beta Decay in the Dualized Standard Model}}\\
 
\vspace{.5cm}

\baselineskip 14pt

{\large Jos\'e BORDES\footnote{Also at IFIC, Centro Mixto Universitat de 
Val\'encia-CSIC}}\\
 jose.m.bordes\,@\,uv.es\\
{\it Departament Fisica Teorica, Universitat de Valencia,\\
  calle Dr. Moliner 50, E-46100 Burjassot (Valencia), Spain}\\

\vspace{.2cm}

{\large CHAN Hong-Mo}\\
chanhm\,@\,v2.rl.ac.uk  \,\,\,  \\
{\it Rutherford Appleton Laboratory,\\
  Chilton, Didcot, Oxon, OX11 0QX, United Kingdom}\\

\vspace{.2cm}

{\large Ricardo GALLEGO}\\
gallegor\,@\,titan.ific.uv.es \\
{\it Departament Fisica Teorica, Universitat de Valencia,\\
  calle Dr. Moliner 50, E-46100 Burjassot (Valencia), Spain}\\

\vspace{.2cm}

{\large TSOU Sheung Tsun}\\
tsou\,@\,maths.ox.ac.uk\\
{\it Mathematical Institute, University of Oxford,\\
  24-29 St. Giles', Oxford, OX1 3LB, United Kingdom}

\end{center}

\vspace{.3cm}

\begin{abstract}

The Dualized Standard Model offers a {\it raison d'\^etre} for 3 fermion
generations and an explanation for their distinctive mass and mixing 
patterns, reproducing to a reasonable accuracy the empirical mixing 
matrix and mass spectrum for both quarks and leptons in terms of just a 
few parameters.  With its parameters thus fixed, the result is a highly 
predictive framework.  In particular, it is shown that it gives explicit
parameter-free predictions for neutrinoless double beta decays.  For 
$^{76}Ge$, it predicts a half-life of $10^{31}-10^{32}$ years, which
satisfies the present experimental lower bound of $1.8 \times 10^{25}$ 
years.

\end{abstract}

\end{titlepage}

\clearpage

Neutrinoless double beta decays of the type:
\begin{equation}
A \rightarrow B \, + \, 2 e^-
\label{process}
\end{equation}
in which a nucleus $A = (Z,N)$ converts into a nucleus $B = (Z+2,N-2)$ 
emitting $2e^-$ with no accompanying neutrinos, has long been recognized as 
a most promising probe for possible lepton-number violation \cite{Furry}.  
It is also the most likely test to decide whether the neutrino is a Majorana 
particle, since the best known possibility for the decay is the exchange 
of such neutrinos.  For this reason, high sensitivity experiments have 
already been done giving very stringent bounds on the decay rate, of which 
the tightest so far is from the Heidelberg-Moscow experiment searching 
for the decay: 
\begin{equation}
^{76}Ge \rightarrow  ^{76}Se + 2 e^-,
\label{GetoSe}
\end{equation}
which after a $31$ kg$\,\cdot\,$year run gives a limit on the half-life:
\cite{HeidelbergMoscow}: 
\begin{equation}
\tau(^{76}Ge) > 1.8 \,\times \, 10^{25} \, \, {\rm years} \, \, \, 
(90 \% \, \, \, {\tt C.L.})
\label{Gedecay}
\end{equation}
Further effort with this experiment, it is claimed, can reach limits of 
up to $10^{27}$ years in one year, and up to $10^{29}$ years in ten years 
of running, so that an improvement of several orders of magnitude is 
foreseeable in the near future.

The decay (\ref{process}) violates lepton-number by 2 units and is thus
forbidden in the conventional version of the Standard Model.  However,
lepton-number is conserved there only by a global symmetry which would 
be broken at some level in almost any extension to the model and would lead 
to neutrinoless double beta decays of the type (\ref{process}).  It is thus 
incumbent upon advocates of any extension to the Standard Model to check 
whether their proposal predicts decay rates for (\ref{process}) which are 
first, consistent with present bounds, and second, 
accessible to future experiment.  

In particular, the Dualized Standard Model (DSM) \cite{dualcons} which we 
ourselves advocate has to be subjected to such a test.  This DSM scheme 
purports to extend the conventional version of the Standard Model in such 
a way as (i) to offer an explanation for the existence of 3 and only 3 
fermion generations, (ii) to deduce the qualitative features of fermion 
mixing and the hierarchical fermion mass spectrum, and (iii) to allow 
a systematic calculation of the mixing parameters both for quarks 
and leptons giving results in general agreement with experiment 
\cite{dualcons,phenodsm}.  However, in the DSM explanation for neutrino 
oscillations, the neutrinos acquire their very small masses through the 
see-saw mechanism \cite{seesaw} with the introduction of right-handed 
singlets.  Lepton-number violation is thus implied giving decays of type
(\ref{process}) so that it is incumbent upon us to check its predictions
in this against experiment.  In fact, as we shall show, the model is so 
constrained by its calculation of the fermion mass and mixing parameters
\cite{phenodsm} that its prediction for (\ref{process}) is now entirely 
parameter-free and explicit.  For (\ref{Gedecay}), in particular, it gives:
\begin{equation}
\tau(^{76}Ge) = 10^{31} - 10^{32} \, \, {\rm years},
\label{Gepredict}
\end{equation}
with the uncertainty coming mostly from the present empirical uncertainty 
in the mass $m_3$ of the heaviest neutrino.  Comparing this result with 
(\ref{Gedecay}), we conclude first, that the DSM scheme survives the present 
empirical bound, which is in itself nontrivial since the prediction is 
parameter-free, and second, that the prediction is between one and two
orders of magnitude below the sensitivity  range of present planned
experiments. 

In what follows, we shall detail how the above result (\ref{Gepredict})
is derived, finishing with a discussion of its possible implications.

First, a few words about the general features of the DSM scheme which lead
to fermion mixing and fermion mass hierarchy.  Using theoretical results 
derived earlier \cite{dualsymm,dualcomm} a candidate for the ``horizontal 
symmetry'' of generations is identified as the dual to colour $SU(3)$ 
\cite{dualcons} which is naturally broken \cite{tHooft} giving 3 and only 
3 generations as a result.  Duality suggests also the mechanism for breaking 
the symmetry \cite{dualcons} leading to fermion mass matrices with only 
one nonzero eigenvalue (rank-one).  At tree-level, this means zero masses 
for the 2 lower generations and no mixing between up- and down- states.  
With loop corrections, however, the mass matrix changes its orientation 
in generation space (rotates) with changing energy scales.  As a result, 
mass ``leaks'' from the heaviest into the 2 lower generations, giving 
the characteristic hierarchical mass spectrum observed.  Further, mixing 
occurs between up and down flavour states, with the mixing matrix elements 
given as direction cosines between the two triads of mass eigenvectors 
at the two scales corresponding to respectively the up- and down-states.  
The framework depends on several parameters related to the vev's and Yukawa 
couplings of the (dual colour) symmetry-breaking Higgs bosons, loops of 
which are what drive the mass matrix rotation.  Of these parameters those 
3 relevant to the mass and mixing patterns were fitted to $m_c/m_t$, 
$m_\mu/m_\tau$, and the Cabibbo angle.  Given then the masses of the 
heaviest generation, one predicts the masses of the other quarks and 
charged leptons together with the remaining quark mixing angles all of 
which are in general agreement with experiment.  For more details the 
reader is referred to e.g. \cite{phenodsm}.

When applying the above mechanism to neutrinos, however, giving just the 
physical mass $m_3$ of the heaviest neutrino is not enough since by the 
see-saw mechanism \cite{seesaw} the physical masses $m_i$ of neutrinos 
depend on both their Dirac masses $M_i$ and the right-handed neutrino 
mass $B$, thus: $m_i = M_i^2/B$, and it is on the Dirac mass matrix that 
mixing depends \cite{nuosc}.  One inputs therefore also the physical mass
$m_2$ of the second generation neutrino, but here, one finds that not all
inputs for $m_2$ will work, for the DSM mechanism, as explained above, 
will give only a hierarchical mass spectrum.  In practical terms, it 
means that of the current 4 admissible solutions for solar neutrinos
\cite{SuperK}, only the so-called vacuum and low solutions are admitted
by DSM.\footnote{We note that although it was thought at one stage
\cite{Takeuchi} that recent data from Superkamiokande exluded both the
SMA and vacuum-low solutions at 95 percent confidence level, later
more thorough anaylses of the global date \cite{concha} show that at
present there is no reason to exlude any of the solutions (LMA, SMA,
vacuum-low) at any reasonable confidence level.}  
Inputting then $m_3^2
\sim 3 \times 10^{-3}\ {\rm eV}^2$ as indicated by atmospheric neutrino data
and $m_2^2 \sim 10^{-10}\ {\rm eV}^2$ for the vacuum solar neutrino solution
\cite{SuperK}, all other required  parameters being already fixed as explained
above, one predicts with DSM  \cite{phenodsm} the lepton mixing matrix
together with all the other  neutrino masses $m_1$, $M_i, i = 1,2,3$, and $B$.
 In particular, one finds  the mixing matrix elements:  \begin{equation}
U_{e 3} \sim 0.07, \ \  U_{\mu 3} \sim 0.66,
\label{Uemu3}
\end{equation}
a right-handed neutrino mass $B$ of order of a few hundred TeV, and the  
mass $m_1$ of the lightest neutrino very small (as low as $10^{-15}$ eV!).
These are all that is needed for our discussion here.  In passing, we note 
that the predicted mixing angles (\ref{Uemu3}) are both in excellent 
agreement with present experiment \cite{SuperK,Chooz}.  We shall return 
later to comment on the accuracy and reliability of the above predictions.
 
Next, turning to neutrinoless double beta decay, an explicit formula for 
the half-life of the process (\ref{process}) has been worked out in full 
generality in, for example, \cite{doikotani} for an effective Lagrangian 
with left and right handed fermionic currents, which formalism we shall 
mainly follow, since the theory involved here is basically contained in 
the Standard Model except for the addition of a right-handed neutrino 
component to allow for lepton number violation.  In figure 1 we give the 
Feynman diagram for the process.  As one can see, the emission of 2 $W$ 
gauge bosons transforms the 2 $d$ quarks into $u$ quarks, and in turn the 
2 neutrons into the corresponding protons.  In each weak vertex an element 
of the MNS mixing matrix \cite{MNS} has to be included, and in the neutrino 
propagator it is the physical mass of the neutrino that enters.  As a result, 
given that the physical neutrino masses are much smaller than the energies 
typically involved, the amplitude is proportional to:
\begin{equation}
\langle m_\nu \rangle = \sum_j m_j U^2_{ej},
\label{effectivemass}
\end{equation}
which may be considered as an effective mass for the neutrino exchanged.

\begin{figure} 
\centerline{\psfig{figure=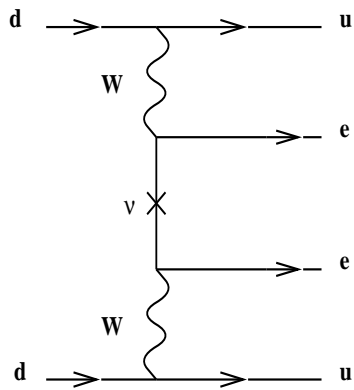,width=.9\textwidth,height=.9\textheight}} 
\vspace{-8cm} 
\caption{Feynmann diagram illustrating the neutrinoless double beta 
decay.} 
\label{Feyndiag} 
\end{figure} 

With these ingredients we have for the inverse half-life of a nucleus in 
a transition between $0^+$ states:
\begin{equation}
\Gamma_{0\nu}(0^+ \rightarrow 0^+)=
\frac{G_F^4 m^5_e}{16 \pi^5} G_{01}(T) g^4_A
| \langle M_{Nucleus} \rangle |^2 \langle m_\nu
\rangle^2. \label{halflife1}
\end{equation}
(We have not shown explicitly the Fermi factor for the emitted electrons 
since it can be consistently substituted by a factor of two.)

The different factors appearing in (\ref{halflife1}) arise as follows. 
First, the Fermi constant raised to the 4th power comes from the two weak 
vertices.  Second, there is the phase space factor $G_{01}(T)$ given by:
\begin{equation}
G_{01}(T)=\frac{1}{15} T (T^4+ 10T^3 + 40 T^2 + 60 T + 30)
\label{G01}
\end{equation}
where $T$ is the maximum kinetic energy attainable by the emitted electrons
in units of the electron mass.  Since for the cases of interest $T$ is in 
the range $(2,3)$, no further simplification in (\ref{G01}) is allowed, but
$T$ itself may be approximated by:
\begin{equation}
T = \frac{M_A- (M_B+2 m_e)}{m_e} \sim 2 \frac{m_n-m_p-m_e}{m_e}.
\label{Tapprox}
\end{equation}
Third, the expectation values for the Hamiltonian in  nuclear states ($\langle
M_{Nucleus}  \rangle$) involving the nuclear structure have been treated 
in the usual effective way, using a non-relativistic approximation for the
nuclear motion and the impulse approximation for the interaction between 
leptons and nucleons.  The nuclear interactions in the limit when the momentum
transfer between nucleons is low compared to the nucleon mass is taken into 
account by three form factors describing the so-called Fermi, Gamow-Teller
and tensor  interactions.
The values of these matrix elements have been calculated for various nuclei in
different nuclear  models \cite{fgt}. The additional factor $g_{A}$ is the
effective form  factor describing the interactions between the electroweak
gauge bosons  and the quarks inside the nucleons at zero momentum transfer
which is  appropriate for this case.  Finally there is the square of the
neutrino  effective mass $\langle m_\nu \rangle$ already mentioned.

One notes that of all the factors appearing in (\ref{halflife1}), only 
the last $\langle m_\nu \rangle^2$ depends specifically on the DSM scheme 
through the neutrino masses and mixing angles.  Given now that in the DSM, 
as explained before, the masses of neutrinos are hierachical, namely that
$m_3 \gg m_2 \gg m_1$ dropping by at least 3 orders of magnitude from
generation to generation, the sum in (\ref{effectivemass}) is dominated
entirely by the $m_3$ term despite the smallness of the MNS mixing element
$U_{e 3}$, giving:
\begin{equation}
\langle m_\nu \rangle \sim m_3 (U_{e3})^2.
\label{effectivemas}
\end{equation}
In other words, to a good approximation, the half-life for neutrinoless
double beta decay depends in the end only on the mass $m_3$ of the heaviest 
neutrino mass eigenstate $\nu_3$ and on the mixing element $U_{e 3}$ from 
the electron neutrino into this state.  The former quantity $m_3$ is known
to a fair accuracy (assuming the masses to be hierarchical) from e.g. the
Superkamiokande experiment \cite{SuperK}, while the latter quantity $U_{e 3}$
was calculated from the DSM scheme giving the value in (\ref{Uemu3}).  The
half-life is therefore explicitly calculable.

In particular, let us apply the formula to the most promising example 
(\ref{GetoSe}) for which a running experiment not only gave already a very 
non-trivial bound but can in the foreseeable future improve considerably 
on the present limits \cite{HeidelbergMoscow}.  We note that in the 
nuclear matrix elements between states of the same spin-parity ($0^+$)
the dominant contribution comes from the Gamow-Teller transition.  This
piece has been evaluated for instance in \cite{fgt} using the pn-RQRPA 
model which gives good results in heavy nuclei, resulting in the value for a
nucleus of $^{76}Ge$  $\langle M_{Ge}\rangle =0.28$ $Gev$. 
Then, with the axial-vector form factor conventionally taken as $g_A=1.24$, we
obtain from (\ref{halflife1}): \begin{equation}
\Gamma_{0\nu}(^{76}Ge)= 4.8 \times 10^{-39} \, \, 
\langle m_\nu \rangle^2 \, \, \, {\rm GeV^{-1}}.
\label{GammaGe}
\end{equation}
Substituting then the DSM value of $U_{e 3}$ from (\ref{Uemu3}) \cite{phenodsm}
and the presently allowed range $m^2_3=10^{-2}-10^{-3}$ eV$^2$ for the 
(physical) mass of the heaviest neutrino \cite{SuperK,numasses}, we obtain: 
\begin{equation}
\Gamma_{0\nu}(^{76}Ge)= (1-10) \times 10^{-64} \, \, \, {\rm GeV}.
\label{GammaGe1}
\end{equation}
Apart from the folding in of some minor numerical uncertainties in the
DSM calculation which we shall now clarify, this corresponds to the range of
half-life for $^{76}Ge$ quoted before in (\ref{Gedecay}).

In the DSM calculation of the mixing parameters (\ref{Uemu3}), besides the
possible uncertainties in the scheme itself which we would not know how to
estimate, the main imprecisions came from the empirical quantities used to
determine the unknown parameters, namely, the quark masses and the 
Cabibbo angle.  This gave a range of values for $U_{e 3} = 0.063 - 0.073$
\cite{phenodsm}, corresponding to a spread in the predicted value of 
$\Gamma_{0\nu}$ only of about a factor 2.  This means that if it were not
for the uncertainties in the nuclear physics and in the present empirical 
value of $m_3$, the prediction of DSM on $\Gamma_{0\nu}$ could be made
much more precise.  The mixing element $U_{e 3}$ depends in principle also
on the ratio $m_2/m_3$, which within the DSM framework is equivalent to
a dependence on the Dirac mass $M_3$ of $\nu_3$, but this dependence is
weak, as demonstrated numerically in \cite{nuosc}, so long as $M_3$ remains
in the MeV range as in the quoted calculation \cite{phenodsm}.  More 
concisely, it can be shown with a little calculation that to a good 
approximation:  
\begin{equation}
U_{e3}=0.15 \sin \left( \theta - \sqrt{\frac{M_2}{M_3}} \right) -
   \frac{1}{2} \sqrt{\frac{M_3}{m_0}},
\label{ue3}
\end{equation}
(where $\theta=\tan^{-1}\sqrt{2}$ and $m_0=1.25$ GeV, which are parameters 
of the model related to the boundary conditions of the renormalization group 
equation, namely the v.e.v's of dual Higgs bosons).  Although the precise 
choice of $M_3$ hardly affects the prediction for the rate of neutrinoless 
double beta decays, we would nevertheless advocate our particular choice 
in \cite{nuosc,phenodsm} of $M_3 \sim 4$~MeV because this corresponds to 
a right-handed neutrino mass $B$ of order 500 TeV, which happens also to 
be the symmetry-breaking scale for the generation (dual colour) symmetry 
in DSM as estimated from FCNC effects \cite{fcnc} and the so-called post-GZK 
cosmic ray air showers \cite{UHECR}.   One is reminded of a parallel 
situation in grand unified theories where it has been widely assumed that 
the mass B is of the same order of magnitude as the unification breaking 
scale, although the actual scale of order $10^{15}$ GeV there is very 
different from the 100~TeV scale advocated here for the DSM.

From eq. (\ref{ue3}), one sees that the mixing element $U_{e3}$ in the 
DSM scheme is generically non-zero.  This can be seen also from the fact 
that in DSM a corner element such as $U_{e3}$ in a mixing matrix (whether 
CKM or MNS) can be interpreted as the effect of torsion of a certain curve 
on the unit sphere which, though vanishing to first order in separation 
between up and down states and therefore small, has no reason to be 
zero for finite separation \cite{features,phenodsm}.  That $U_{e3}$ is
nonzero is what makes the DSM prediction for the rate of neutrinoless
double beta decay non-vanishing, and thus in principle observable
though not perhaps in the foreseeable future, 
and distinguishes it from models with exact
bimaximal mixing  which predicts the element $U_{e 3}$ to be precisely zero,
in which case  the experimental observation of the effect would be 
ruled out.

To advocates of the DSM scheme like ourselves the fact that its prediction 
for neutrinoless double beta decay survives existing bound is of course  
a relief, given that the calculation has no freedom 
and depends on no adjustable parameters.  However, so long as the effect 
remains undiscovered, it is not a positive test for the validity of the 
scheme like the original tests on the fermion mass and mixing parameters 
\cite{phenodsm}.  Still it adds to the list of DSM predictions which have
survived experimental bounds after the parameters of the scheme had been 
fixed.  This now includes FCNC effects in meson mass differences and decays 
\cite{fcnc}, post-GZK cosmic ray airshowers \cite{UHECR}, $\mu-e$ conversions 
in nuclei and muonium \cite{mueconv}, and lepton-flavour violation due to 
``transmutation'' \cite{impromat} in $\gamma e$ \cite{photrans} and $e^+ e^-$ 
\cite{transbha} collisions, and in vector boson decays \cite{transcay}.  
Together now with neutrinoless double beta decay, the total weight even of
just these survival tests is beginning to look somewhat nontrivial.  Moreover,
since some of the effects predicted are well within present experimental 
sensitivity, there is good hope that they may be positively tested in the 
not too distant future. 
 
One of us (J.B.) has been supported in part by grants AEN99/0692, 
PB97-1261 and GV98-1-80.

\end{document}